\documentstyle[twocolumn,pra,aps,epsfig]{revtex}
\begin{document}
\flushbottom
\draft
\title{Theory of superradiant scattering of laser light
from Bose-Einstein condensates}
\author{M. G. Moore and P. Meystre}
\address{Optical Sciences Center and Department of Physics\\
University of Arizona, Tucson, Arizona 85721\\
(May 28, 1999)
\\ \medskip}\author{\small\parbox{14.2cm}{\small \hspace*{3mm}
In a recent MIT experiment, a new form of superradiant Rayleigh scattering
was observed in Bose-Einstein condensates. 
We present a detailed theory of this phenomena in which  
condensate depeletion leads to mode competition, which, together with
the directional dependence of the scattering rate,
is ultimately responsible for the observed phenomena. The nonlinear response 
of the system is shown to be highly sensitive
to initial quantum fluctuations which cause large run to run
variations in the observed superradiant pulses.
\\[3pt]PACS numbers: 03.75.-b,03.75.Fi,42.50.Fx  }}
\maketitle
\narrowtext

With the recent advent of Bose-Einstein condensation (BEC)
in low-density alkali vapors \cite{AndEnsMat95,DavMewAnd95}, a laser-like
source of coherent monochromatic atomic matter-waves is now readily
available. As the electromagnetic vacuum itself provides a nonlinear medium
for atomic fields, an atomic BEC is thus an ideal system to study nonlinear 
wave-mixing and related phenomena.  Indeed, 
nonlinear atom optics \cite{LenMeyWri93,ZhaWalSan94,CasMol95} is now an 
experimental reality with the recent observation of atomic four-wave mixing in 
condensate systems
\cite{SteInoSta98,Phi4wm}. 
In addition to wave-mixing between atomic matter waves, the 
ability to generate
laser-like atomic fields also raises the possibility to observe direct
wave-mixing between atomic and optical fields. 

In the case where the atoms interact only with far off-resonant optical 
fields, the 
dominant atom-photon interaction is two-photon Rayleigh 
scattering.  When the atoms are described as matter waves, 
Rayleigh scattering is formally equivalent to a cubic nonlinearity, 
and therefore leads to 
four-wave mixing between atomic and optical fields. In recent
work\cite{MooMey99a,MooMey99b,LawBig98} along these lines, 
the scattering of light by a condensate from a strong pump laser into
a weak quantized optical cavity mode was considered. This work was an 
extension of the collective atomic recoil laser (CARL) 
\cite{BonSalNar94,MooMey98} into the regime of BEC,
and focused on exploiting an instability in the light-matter
interaction to parametrically amplify atomic and optical waves as well as 
to optically manipulate matter-wave coherence properties and generate 
entanglement between atomic and optical fields. 

Recent experiments by Ketterle and coworkers ar MIT 
\cite{InoChiSta99}, however, have demonstrated that
this instability can play an important role also in the case in which laser 
light is scattered into the vacuum modes of the electromagnetic field. In
these experiments, a variation of Dicke superradiance \cite{Dic54} was observed 
in which the role of electronic coherence, which stores the memory
of previous scattering events, is replaced by coherence
between center-of-mass momentum states, i.e. interference fringes in the atomic
density. In this paper we present a multi-mode theory of condensate
superradiance. Beginning with the 
elimination of the radiated light field as in the Wigner-Weiskopf 
theory of spontaneous emission, we then derive a linearized model which
describes amplification of quantum fluctuations. This is then coupled to a
`classical' nonlinear model in which 
mode competition quenches
scattering in all but the direction(s) of maximum gain. The initial 
quantum fluctuations are shown to strongly influence superradiant pulse 
formation, and lead to large fluctuations between runs with identical 
experimental parameters. In the MIT experiments this is clearly demonstrated by
the presence of random spots in the angular distribution of the scattered
photons. 

Our model consists of a Schr\"odinger field of two-level atoms  
coupled via the
electric-dipole interaction to a far-off resonant pump laser field,
as well as to the vacuum modes of the electromagnetic field.
The pump laser has frequency $\omega_0$, wavevector ${\bf
k}_0=(\omega_0/c)\hat{\bf y}$, and its polarization is taken along
the $\hat{\bf x}$ axis. 
Due to the large detuning between the pump frequency 
and the atomic transition frequency $\omega_a$, we can
eliminate the excited state field, and describe the atoms
as a scalar field of ground state atoms. This atomic field is self-interacting,
due to ground-ground collisions, however,
we note that collisions which transfer populations are
generally nonresonant and should make only a small contribution to the dynamics.
The remaining collisions then simply give a mean field
shift to the resonance frequency for quasiparticle excitations.  
As the effect of these shifts on our model are negligible, at
present we include collisions only implicity in the determination of the
condensate wavefunction.

The effective Hamiltonian which describes the coupling of
the atomic and electromagnetic fields is given by
\begin{eqnarray}
\hat{H}&=&\int d^3{\bf r}\,\hat{\mit\Psi}^\dag({\bf r})H_0({\bf r})
\hat{\mit\Psi}({\bf r})+\int d^3{\bf k}\, 
\hbar\omega({\bf k})\hat{b}^\dag({\bf k})
\hat{b}({\bf k})\nonumber\\
&+&\int d^3 {\bf k}\,d^3{\bf r}\,
\big[\hbar g({\bf k})\hat{\mit\Psi}^\dag({\bf r})\hat{b}^\dag({\bf k})
e^{i({\bf k}_0-{\bf k})\cdot{\bf r}}\hat{\mit\Psi}({\bf r})\nonumber\\
&+&H.c.\big]
\label{H1}
\end{eqnarray}
where $\hat{\mit\Psi}({\bf r})$ is the atomic field operator, and $\hat{b}({\bf
k})$ is the annihilation operator for a photon in mode ${\bf k}$ in the frame
rotating at the pump frequency $\omega_0$. The photon energy in this frame is 
given by $\omega({\bf k}) =c|{\bf k}|-\omega_0$.
The single-atom Hamiltonian is given by $H_0({\bf r})=-(\hbar^2/2m)\nabla^2+V(r)
+\hbar|\Omega_0|^2/2\Delta$, $V({\bf r})$ being the trap potential, 
$\Omega_0$
the pump Rabi frequency, and $\Delta=\omega_0-\omega_a$ the
pump detuning. 
The Hamiltonian (\ref{H1}) includes only scattering 
of pump photons, i.e.
multiple scatterings between vacuum modes are neglected. The coupling
coefficient for Rayleigh scattering between the pump and vacuum modes is 
\begin{equation}
g({\bf k})=\frac{|\Omega_0|}{2|\Delta|}
\sqrt{\frac{c|{\bf k}|d^2}{2\hbar\epsilon_0(2\pi)^3}}|\hat{\bf k}\times\hat{\bf x}|,
\label{defg}
\end{equation}
where $d$ is the magnitude of the atomic dipole moment for the transition
involved. 

The atomic field is initially taken to be a number state
in which $N$ atoms occupy the trap ground state $\varphi_0({\bf r})$, which 
satisfies $[H_0({\bf r})-\hbar\mu]\varphi_0({\bf r})=0$, $\mu$ being the 
energy of the trap ground state.
The effect of atomic recoil during Rayleigh scattering between the
pump and the vacuum mode ${\bf k}$ is therefore to transfer atoms 
into the state $\varphi_0({\bf r})\exp[i({\bf k}_0-{\bf k})\cdot{\bf r}]$.
This suggest to expand the atomic field operator onto quasi-modes
according to 
\begin{equation}
\hat{\mit\Psi}({\bf r},t)=\sum_{\bf q}\langle{\bf r}|{\bf q}\rangle
e^{-i(\omega_{\bf q}+\mu)t}\hat{c}_{\bf q}(t),
\label{modes}
\end{equation}
where $\langle{\bf r}|{\bf q}\rangle=\varphi_0({\bf r})
\exp\left(i{\bf q}\cdot{\bf r}\right)$, and
$\omega_{\bf q}=\hbar|{\bf q}|^2/2m$.
This is similar to the slowly varying envelope approximation from optical
physics, the envelope being given here by $\varphi_0({\bf r})$.

A discrete quantization of the ${\bf q}$ values follows from
the requirement that the operators $\{ \hat{c}_{\bf q}\}$ 
obey boson commutation relations $[\hat{c}_{\bf q},
\hat{c}^\dag_{{\bf q}^\prime}]=\langle{\bf q}|{\bf q}^\prime\rangle\approx
\delta_{{\bf q},{\bf q}^\prime}$. Due to the finite size of the ground state 
wavefunction $\varphi_0({\bf r})$, this means that ${\bf q}$ and 
${\bf q}^\prime$ must be seperated in ${\bf k}$-space.  
Hence, the summation in Eq. (\ref{modes}) is taken to include the condensate 
mode ${\bf q}=0$ as 
well as a grid of ${\bf q}$-values as closely spaced as is  
consistent with orthogonaliy. Clearly,
this expansion is not rigorously orthogonal and complete,
however, it is sufficient to account for the quantum statistical effects which
occur above the critical phase-space density. 

An important aspect of BEC superradiance is the generation of 
families of higher-order sidemodes due to the scattering of pump photons by
the first-order sidemodes. For the scope of this paper, however, we
consider a simplified model containing only the primary Rayleigh scattering 
process whereby a condensate atom is transferred to a
first-order sidemode by scattering a pump photon. With this simplification,
we insert the expansion (\ref{modes}) into Eq. (\ref{H1}) and arrive at the
effective Hamiltonian 
\begin{eqnarray}
\hat{H}&=&\int d^3{\bf k}\,\hbar\omega({\bf k})\hat{b}^\dag({\bf k})
\hat{b}({\bf k})\nonumber\\
&+&\sum_{{\bf q}\ne 0}\int d^3{\bf k}\,
\left[\hbar g({\bf k})\rho_{\bf q}({\bf k})e^{i\omega_{\bf q}t}
\hat{c}^\dag_{\bf q}\hat{b}^\dag({\bf k})\hat{c}_0+H.c.\right],
\label{H2}
\end{eqnarray}
where $\rho_{\bf q}({\bf k})=\int d^3{\bf r}\, |\varphi_0({\bf r})|^2
\exp[-i({\bf k}-{\bf k}_0+{\bf q})\cdot{\bf r}]$
is the Fourier transform
of the ground state density distribution
centered at ${\bf k}={\bf k}_0-{\bf q}$. 

From the Hamiltonian (\ref{H2}) it is straightforward to derive the
equation of motion for $\hat{b}({\bf k})$, which upon formal integration
yields
\begin{eqnarray}
\hat{b}({\bf k},t)&=&\hat{b}({\bf k},0)e^{-i\omega({\bf k})t}
-i\sum_{{\bf q}\ne 0} g({\bf k})\rho_{\bf q}({\bf k})e^{i\omega_{\bf q}t}
\nonumber\\
&\times&\int^t_0d\tau e^{-i(\omega({\bf k})+\omega_{\bf q})\tau}
\hat{c}^\dag_{\bf q}(t-\tau)\hat{c}_0(t-\tau),
\label{b}
\end{eqnarray}  
where the first term gives the free electromagnetic field, i.e.
vacuum fluctuations, and
the second term is the radiation field due to Rayleigh scattering.
A nonzero expectation value of the coherence operator 
$\hat{c}^\dag_{\bf q}\hat{c}_0$ indicates the presence of interference 
fringes, hence the radiated field increases as fringes build 
up. {\it This term therefore leads to an instability where the memory of 
previous scattering events, stored in the matter-wave interference 
fringes, enhances the present rate of Rayleigh scattering}.

Equation (\ref{b}) is then substituted into the equation of motion
for $\hat{c}_{\bf q}$.In the Markoff approximation, familiar
from the Wigner-Weisskopf theory of spontaneous emission, this yields
\begin{eqnarray}
\frac{d}{dt}\hat{c}_{\bf q}&=&
-i\int d^3{\bf k}\, g({\bf k})\rho_{\bf q}({\bf k})
\hat{b}^\dag({\bf k},0)
e^{i(\omega({\bf k})+\omega_{\bf q})t}\hat{c}_0\nonumber\\
&+&\frac{G_{\bf q}}{2}
\hat{c}^\dag_0\hat{c}_0\hat{c}_{\bf q},
\label{dcqdt}
\end{eqnarray}
where
\begin{equation}
G_{\bf q}=2\pi\int d^3{\bf k}\, |g({\bf k})|^2|\rho_{\bf q}({\bf k})|^2
\delta(\omega({\bf k})+\omega_{\bf q})
\label{G}
\end{equation}
is the single-atom gain. In deriving Eq. (\ref{dcqdt}) we have used
the orthogonality of the states $\{|{\bf q}\rangle\}$
to make the approximation $\rho^\ast_{\bf q}({\bf k})
\rho_{{\bf q}^\prime}({\bf k})
\approx|\rho_{\bf q}({\bf k})|^2\delta_{{\bf q},{\bf q}^\prime}$,
and neglected the principal
part which accompanies the $\delta$-function. If included, the principle part
would contribute additional ground-state collisions due to the  
dipole-dipole interaction. 

For a closed atomic system, the total number of atoms is conserved, hence
$\hat{c}^\dag_0\hat{c}_0=N-\sum_{{\bf q}\ne 0}\hat{c}^\dag_{\bf q}
\hat{c}_{\bf q}$. For very short times we can therefore take
$\hat{c}^\dag_0\hat{c}_0\approx N$.
In this case Eq. (\ref{dcqdt}) reduces to
\begin{equation}
\frac{d}{dt}\hat{c}_{\bf q}=\frac{G_{\bf q}}{2}N\hat{c}_{\bf q}+\hat{f}^\dag_{\bf q}(t).
\label{linear}
\end{equation}
where 
$\hat{f}_{\bf q}(t)$ is a noise operator whose correlation functions
are given in the Markoff approximation by
\begin{eqnarray}
\langle\hat{f}^\dag_{\bf q}(t)\hat{f}_{\bf q}(t^\prime)\rangle&=&0,\nonumber\\
\langle\hat{f}_{\bf q}(t)\hat{f}^\dag_{\bf q}(t^\prime)\rangle
&=&G_{\bf q}N\delta(t-t^\prime).
\label{noise}
\end{eqnarray}
These noise operators allow the system to be triggered by quantum fluctuations,
and hence describe ``spontaneous'' scattering which occurs in the
absence of any sidemode population.

Equation (\ref{linear}) can be solved exactly, giving
\begin{equation}
\hat{c}_{\bf q}(t)=e^{(G_{\bf q}/2)Nt}\hat{c}_{\bf q}(0)
+\int^t_0d\tau\, e^{(G_{\bf q}/2)N\tau}\hat{f}^\dag_{\bf q}(t-\tau).
\label{quantum}
\end{equation}
From Eq. (\ref{quantum}) it is possible to compute 
the probability $P_{\bf q}(n,t)$
of having $n$ atoms in mode ${\bf q}$ at time $t$, assuming zero population at
$t=0$. To accomplish this we
first compute the anti-normally ordered characteristic function,
$\chi_{\bf q}(\eta)=\langle \exp[-\eta^\ast\hat{c}_{\bf q}]
\exp[\eta\hat{c}^\dag_{\bf q}]\rangle$, yielding
\begin{equation}
\chi_{\bf q}(\eta)=e^{-|\eta|^2(\bar{n}_{\bf q}(t)+1)},
\label{chi}
\end{equation}
where $\bar{n}_{\bf q}(t)=\exp(G_{\bf q}Nt)-1$
is the mean population of mode
${\bf q}$ at time $t$. We can identify expression (\ref{chi}) as corresponding
to a chaotic field \cite{WalMil94}. 
The number distribution for a chaotic field is
given by 
\begin{equation}
P_{\bf q}(n,t)=\frac{1}{\bar{n}_{\bf q}(t)}
\left(1+\frac{1}{\bar{n}_{\bf q}(t)}\right)^{-(n+1)},
\label{Pn}
\end{equation}  
which for $\bar{n}_{\bf q}(t)\gg 1$ is well approximated by
$\exp[-n/\bar{n}_{\bf q}(t)]/\bar{n}_{\bf q}(t)$.

When the mean population of a field mode 
is sufficiently large, correlation functions
effectively factorize to all orders, and it becomes possible to formulate a 
`classical' description of the field dynamics.
In the classical theory,
we can consider the sidemode populations as $c$-numbers
and neglect the influence of the quantum noise operators as
spontaneous scattering of atoms is now negligible in comparison to
the stimulated contribution. With the inclusion of condensate depletion,
the sidemode populations then obey the equations
\begin{equation}
\frac{d}{dt}n_{\bf q}=G_{\bf q}(N-\sum_{{\bf q}^\prime\ne 0}n_{{\bf q}^\prime})
n_{\bf q}.
\label{nonlinear}
\end{equation}
Momentum conservation tells us that for each atom
scattered into the sidemode ${\bf q}$, there is a photon 
scattered roughly in the
direction ${\bf k}={\bf k}_0-{\bf q}$. Hence, $N\times dI_{\bf q}/dt$ is the 
ideal photon count rate  
generated by the $|0\rangle\to|{\bf
q}\rangle$ atomic center-of-mass transition. 

The `classical' nonlinear model is applicable when 
$\bar{n}_{\bf q}(t)\gg 1$, whereas the linearized quantum theory requires
$\sum_{{\bf q}\ne 0}\bar{n}_{\bf q}(t)\ll N$.
Provided that the number of active modes is not comparable to $N$,
there is a significant overlap in the validity regimes of these two models.
We then join them by choosing initial conditions for
Eqs. (\ref{nonlinear}) from $P_{\bf q}(n,t_{\mbox{cl}})$, where
$t_{\mbox{cl}}$ satisfies $1\ll \bar{n}_{\bf q}(t_{\mbox{cl}})\ll N$. 
Because the response is still linear at time $t_{\mbox{cl}}$
the resulting nonlinear evolution does not depend on the particular choice
of $t_{\mbox{cl}}$. 

We now analyze the geometrical dependence of the single-atom gain
given by Eq. (\ref{G}), and show that it is largest for radiation along the
long-axis of the condensate. We first
note that $G_{\bf q}$ depends on $g({\bf k})$, which
contains the dipole radiation pattern, as well as on
$\rho_{\bf q}({\bf k})$, which depends on the geometry of
the initial condensate. For a cigar-shaped condensate, aligned
along the $\hat{\bf z}$-axis, $\rho_{\bf q}({\bf k})$ is a disc 
which lies parallel to the $\hat{\bf x}$-$\hat{\bf y}$-plane in
${\bf k}$-space. The dimensions of the disc in ${\bf k}$-space are  
roughly the inverse of the condensate dimensions in ${\bf r}$-space. 
Thus for a condensate
whose dimensions are large compared to an optical wavelength, the
dimensions of $\rho_{\bf q}({\bf k})$ are small compared to $k_0$.

Since $g({\bf k})$ is slowly varying compared to $\rho_{\bf q}({\bf k})$, 
it can be removed from the integral in Eq. (\ref{G}), and evaluated at the
center of $\rho_{\bf q}({\bf k})$. In addition we neglect the
recoil shift $\omega_{\bf q}$ in the $\delta$-function as it has negligible 
effect on the value of $G_{\bf q}$. The remaining integral then
defines the solid angle $\Omega_{\bf q}$ for the scattered radiation associated 
with the ${\bf q}$th mode according to
\begin{equation}
\Omega_{\bf q}=\frac{1}{k_0^2}\int d^3{\bf k}\,
|\rho_{\bf q}({\bf k})|^2\delta(|{\bf k}|-k_0),
\label{Omega}
\end{equation}
which shows that only ${\bf q}$ values for which the center of 
$\rho_{\bf q}({\bf k})$ lies at a distance $k_0$ from the origin
experience gain, a consequence of energy conservation.
Thus for every active 
quasi-mode ${\bf q}$ there is a corresponding radiation direction
$\hat{\bf k}$, such that ${\bf q}=k_0(\hat{\bf y}-\hat{\bf k})$. 

We can obtain a good estimate for $\Omega_{\bf q}$ by taking
$|\rho_{\bf q}({\bf k})|^2$ to be an ellipsoid solid with the inverse
dimensions of the condensate. This gives 
\begin{equation}
\Omega_{\bf q}=\frac{4\pi}{k^2_0 W^2}
\left[\cos^2\theta_{\hat{\bf k},\hat{\bf z}}
+(L/W)^2\sin^2\theta_{\hat{\bf k},\hat{\bf z}}\right]^{-1/2},
\label{Omega2}
\end{equation}
where $L$ is the length of the 
condensate along the $\hat{\bf z}$ axis, $W$ is the radial diameter,
and $\theta_{\hat{\bf k},\hat{\bf z}}$ is the 
angle between the
radiation direction and the long-axis of the condensate. 
Thus $\Omega_{\bf q}$ is maximized for $\theta_{\hat{\bf k},\hat{\bf z}}=0,\pi$,
corresponding to
radiation along $\hat{\bf z}$ and $-\hat{\bf z}$, where
it is given by $\Omega_{\bf q}=4\pi/k_0^2 W^2$.
As $\theta_{\hat{\bf k},\hat{\bf z}}$ moves away from the $\hat{\bf z}$ axis,
$\Omega_{\bf q}$ is relatively flat until it reaches the geometric angle
$W/L$, after which it drops off rapidly.
It is important to note that for the isotropic case $L=W$ there is no preferred 
direction, and a ring of radiation is instead observed.  

Taking into account all of these considerations, 
the expression for the single-atom gain becomes
\begin{equation}
G_{\bf q}={\cal G}\frac{\sin^2\theta_{\hat{\bf k},\hat{\bf x}}}
{\sqrt{\cos^2\theta_{\hat{\bf k},\hat{\bf z}}
+(L/W)^2\sin^2\theta_{\hat{\bf k},\hat{\bf z}}}},
\label{G2}
\end{equation}
where ${\cal G}=3|\Omega_0|^2\Gamma/8|\Delta|^2k_0^2W^2$ is the maximum
single-atom gain, $\Gamma=k_0^3d^2/3\pi\hbar\epsilon_0$ being the
single-atom spontaneous decay rate, and $\theta_{\hat{\bf k},\hat{\bf x}}$
is the angle between the radiation and
polarization directions. 
For the parameters of the MIT experiment
we find ${\cal G}\sim 4\times 10^{-4}\cdot I$, where
$I$ is the laser intensity in mW/cm$^2$, and ${\cal G}$ is given in Hz. 
A rough estimate of the
duration of a superradiant pulse for the case $N=10^6$ and
$I=100$ mW/cm$^2$ is 
$t=\ln(N)/{\cal G}N\sim 150$ $\mu$s, in excellent agreement with
experimentally observed time scales. 
  
The interplay between the dependence of $G_{\bf q}$ on the
radiation direction and the nonlinearity in Eq. (\ref{nonlinear}) 
leads to mode competition, the outcome of which depends
sensitively on the initial quantum fluctuations. When modes with different
values of $G_{\bf q}$ compete, the competition is `unfair'
and the mode with the largest $G_{\bf q}$ generally depletes all of the 
condensate atoms before the populations of the other modes have a chance to 
grow. Modes with the same $G_{\bf q}$, such as
the quasi-modes corresponding to radiation along
the $\hat{\bf z}$ and $-\hat{\bf z}$ directions, instead compete `fairly',
and while the interplay is highly sensitive to
the random initial conditions, neither mode necessarily wins, i.e. there
is no real `winner-takes-all' effect, and often a `tie' will occur.   

The angular dependence of the scattered light from a typical simulation
is shown in Fig. 1, where we have plotted the photon count density
as recorded by an ideal detector array located at a distance $Z$ from the
center of the condensate along the $\hat{\bf z}$ axis. The dark regions
correspond to maximum light intensity, while the white regions indicate
negligible intensity. The center of the figure lies along the symmetry axis of
the BEC, and half-width of the box, given by $ZW/L$, corresponds to the
geometric radiation angle. The simulation was performed for a condensate
with $N=10^6$ atoms in a BEC with a width of 10 $\mu$m and a length of 
100 $\mu$m, in
rough agreement with the MIT experiment. Only  
matter-wave quasimodes corresponding to radiation within twice the 
geometric angle of $+\hat{\bf z}$ and $-\hat{\bf z}$ were included in
the simulation. This set of $576$ modes, is more than sufficient, however, 
as a buildup of significant population in modes corresponding to radiation 
outside of the geometric angle was never observed.
This can be attributed to the fact that these modes have a much 
smaller gain, and thus compete `unfairly' with the 
`endfire' modes. Within $\theta_{\hat{\bf k},\hat{\bf
z}}<W/L$ we see the results of the interplay between quantum fluctuations and
`fair' mode competition. The pattern of dark spots indicates 
multimode superradiance, and exhibits variation on the scale of  
$Z/k_0W$, corresponding to the solid angle of radiation for
an endfire mode. The pattern, which arises from the amplification of quantum
fluctuations, varies randomly from run to run, an effect which has been directly
observed experimentally. 

In conclusion, we note that the quasi-mode populations will experience losses 
as the recoiling atoms eventually
propagate out of the condensate volume.
The lifetime of the quasi-mode, however, is on the order of 
$T_{\bf q}\equiv mL_{\bf q}/\hbar|{\bf q}|$, where $L_{\bf q}$ is the 
length of the condensate along ${\bf q}$. These losses tend to destroy
the coherence between condensate and quasi-mode, which accounts
for the observation of a threshold for 
superradiance in the MIT experiment: for insufficient laser power, the
growth of matter-wave coherence cannot overcome the losses.
As this threshold is very small, we have considered only the situation far above
threshold, in which case the losses may be neglected.

In general, the rate of matter-wave decoherence in 
superradiance or CARL type experiments is given by the ratio between 
the recoil velocity and the matter-wave 
\centerline{\psfig{figure=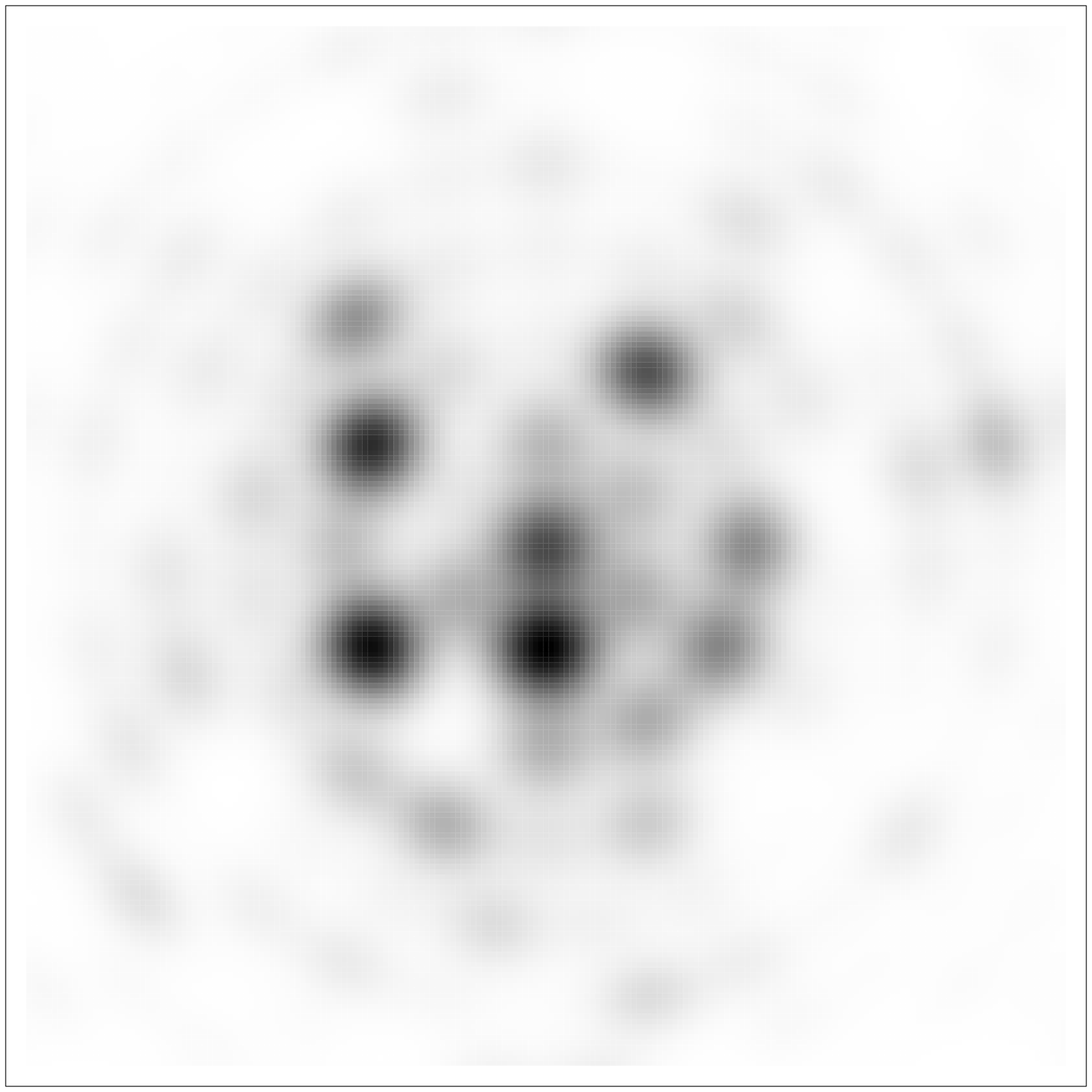,width=7cm,clip=1}}
\begin{figure}
\caption{A typical numerical simulation of condensate superradiance. The 
scattered photon intensity is plotted as seen by a detector array located at a 
distance $Z$ along the symmetry axis of the cigar-shaped BEC.
The blacg regions correspond to maximum, and the white regions
to negligible intensity, respectively. The width of the figure is $2ZW/L$, 
corresponding to radiation within the geometric angle of the condensate.}
\end{figure}
\noindent coherence
length. As a BEC is maximally coherent,
its coherence length is given, as above, by its spatial extent. For a 
noncondensed atomic cloud, however, it is instead given by the thermal DeBroglie 
wavelength,
which is in general small compared to that of the BEC. 
As a result, the threshold for superradiance 
is significantly larger,for which reason superradiance was not observed above 
$T_c$ in the MIT experiment. We remark that in the case of the CARL, 
the presence of an optical 
cavity provides additional feedback, which can compensate for the
lack of atomic coherence and allow for instability and gain. 

Lastly, we remark that the Hamiltonian (\ref{H1}) describes the creation
of correlated atom-photon pairs, and is therefore analogous to the optical
parametric amplifier (OPA), which generates entangled two-photon
states for a variety of applications, e.g. tests of Bells inequality, 
quantum cryptography, and quantum teleportation.
It should therefore be possible to perform analogous experiments using
entangled atom-photon states generated in a BEC superradiance
experiment. 
 
We would like to thank W. Ketterle 
for providing us with an early draft of his experimental results.
This work is supported in part by Office of Naval Research Research Contract 
No. 14-91-J1205, National Science Foundation Grant PHY98-01099, the Army
Research Office and the Joint Services Optics Program.

\end{document}